\documentclass[a4paper,12pt]{article}
\usepackage{a4,graphicx,overcite}

\renewcommand{\baselinestretch}{1.2}

\newcommand{\arXiv}[2][quant-ph]{%
eprint arXiv:\linebreak[0]#1/\linebreak[0]#2%
}

\newcommand{\pra}{Phys.\ Rev.\ A\ }
\newcommand{\prl}{Phys.\ Rev.\ Lett.\ }

\begin{document}

\begin{center}

\textbf{\large How well can you know the edge of a quantum pyramid?} 

\bigskip

Dagomir Kaszlikowski$^*$,
Ajay Gopinathan$^{*,\dagger}$,
Yeong Cherng Liang$^*$,\\
Leong Chuan Kwek$^{\dagger,*}$,
Berthold-Georg Englert$^*$\\

\medskip

$^*$\textit{Quantum Information Technology Group, Department of Physics,\\
  National University of Singapore, Singapore 117542}\\
$^\dagger$\textit{National Institute of Education,
Nanyang Technological University,\\ Singapore 637616}

\bigskip

(10 July 2003)

\bigskip

\renewcommand{\baselinestretch}{1.1}\small
\textbf{Abstract}

\medskip

\begin{minipage}{0.9\textwidth}\small
The methods of quantum cryptography enable one to have
perfectly secure communication lines, whereby the laws of quantum physics
protect the privacy of the data exchanged. 
Each quantum-cryptography scheme has its own security criteria that need to
be met in a practical implementation.
We find, however, that the generally accepted criteria are flawed for a whole
class of such schemes.
\end{minipage}

\renewcommand{\baselinestretch}{1.2}\normalsize

\end{center}

\bigskip\bigskip

\noindent%
Quantum states that, geometrically speaking, constitute the edges of a
high-dimensional pyramid are such that there are equal transition
amplitudes between each pair of states.
Then, if either ``edge state'' occurs with uniform \emph{a priori} 
probability, one cannot be sure which one is the case.
The odds for guessing the state right are maximized by a well-known
measurement scheme, the ``square root measurement'' (SRM) \cite{SRM}, 
but, as we show here, optimal knowledge---in the
information-theoretic sense---is obtained by a different procedure, the
``information maximizing scheme'' (IMS).
Our findings may be of considerable importance for the security of quantum
cryptography and also seem to have a bearing on the quest for optimal quantum 
cloning machines. 

In the situations relevant for quantum cryptography, the common angle between
the edges of the state pyramid is acute.%
\footnote{See, e.g., ref.~\citen{Liang+4:03} and references therein, in particular 
refs.~\citen{Bruss+1:02} and \citen{Acin+2:03}}\  
The SRM is essentially a projection onto one of the edges of a related
pyramid with right angles between the edges.
In marked contrast, the IMS projects either onto the edges of a pyramid 
with an obtuse angle between the edges or onto the
symmetry axis of the state pyramid.
The case of a three-dimensional pyramid is illustrated in the insert 
of Fig.~1.
Matters are analogous in $N=4,5,\dots$ dimensions. 

\renewcommand{\baselinestretch}{1.0}\small

\begin{figure}[t]
\centering\includegraphics[width=0.9\textwidth]{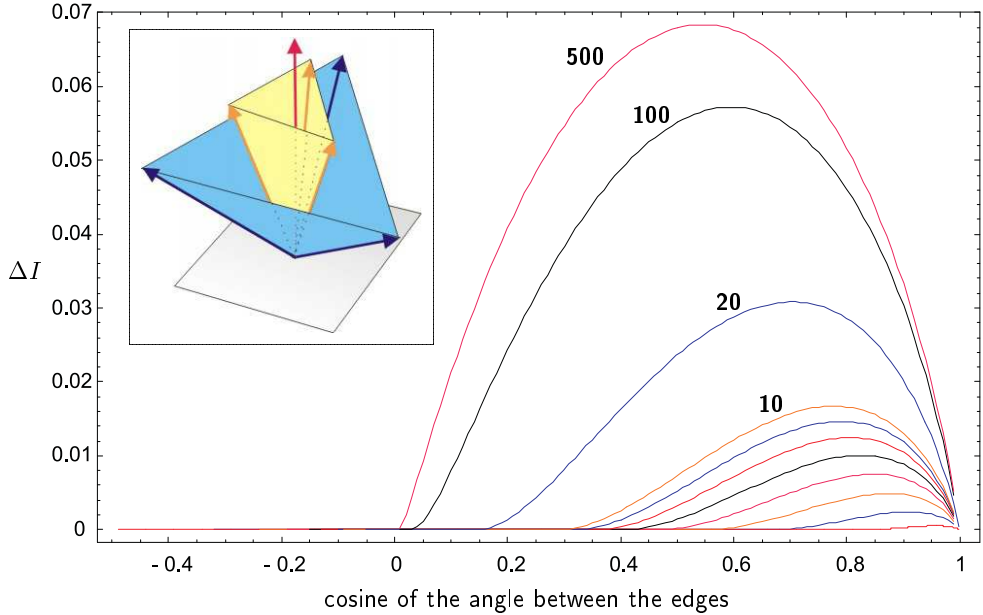}
  
\caption{Difference $\Delta I=I_\mathrm{IMS}-I_\mathrm{SRM}$ 
in the mutual information as obtained by the IMS or the SRM, 
for $N=3,\dots,10,20,100,500$, as a function of the cosine of the common angle
between the edges.
The insert illustrates the case of three-dimensional pyramids: In the yellow
state pyramid the common angle is acute. For the IMS, one has the blue
pyramid with an obtuse angle between the edges, plus a projection on the
state that corresponds to the red arrow,
the symmetry axis of the yellow state pyramid. For the SRM, there is no such
red-arrow contribution and the respective blue pyramid has pairwise
perpendicular edges.}
\end{figure}

\renewcommand{\baselinestretch}{1.2}\normalsize

Figure~1 shows by how much the IMS outperforms the SRM for 
$N=3,4$, $\dots, 10, 20, 100, 500$ by plotting,
as a function of the cosine of the angle between the edges,
the difference 
$\Delta I=I_\mathrm{IMS}-I_\mathrm{SRM}$ in the 
(properly normalized\footnote{That is: All Shannon entropies are computed with
  logarithms to base $N$.})
mutual information 
that would be shared between the sender and receiver of the edge states 
of the state pyramid.
In the quantum-cryptography application, the mutual information values
$I_\mathrm{IMS}$ and $I_\mathrm{SRM}$ are the relevant
numerical measures of what the eavesdropper can find out by the respective
measurement schemes.
At least for the $N$ range in the figure, 
we note that (i) $\Delta I$ increases with the number $N$ 
of transmitted states;
and (ii) that the range of angles, for which this difference is substantial,
also grows with $N$.
For pyramids in this angle range, the usual security analysis, which considers
only the SRM scheme, is thus invalid and must be modified to account for the
better performance of the IMS.

In Fig.~1 we are focusing on the narrow pyramids of cryptographic
relevance. 
For $N=3$, Shor~\cite{Shor:02} found a similar deviation 
for rather wide pyramids (negative abscissa values in the plot), but they
are of no concern here.

In the higher-dimensional quantum cryptography protocols analyzed so far
\cite{Bruss+1:02,Acin+2:03,Liang+4:03} 
the eavesdropper examines state pyramids of the kind discussed
here and performs the SRM.
Our results thus affect the security criteria for these protocols 
in favor of the eavesdropper as she can find out more by employing 
the IMS rather than the SRM. 
We note further that the security analysis in 
refs.~\citen{Bruss+1:02} and \citen{Acin+2:03}, 
which is actually based on universal cloning machines,
is equivalent to considering states pyramids
with the SRM. 
Accordingly, our finding that the IMS is superior 
implies that one must reexamine the criteria 
by which one judges the optimality of universal cloning machines in
applications to quantum cryptography.

\renewcommand{\baselinestretch}{1.0}

\end{document}